\documentclass[12pt]{article}
\usepackage{amsmath,amssymb}
\usepackage{graphicx}
\usepackage{epstopdf}
\usepackage[pdftex]{hyperref}
\DeclareGraphicsExtensions{.eps,.bmp,.wmf,.jpeg,.pdf}
\numberwithin{equation}{section}
\def\be{\begin{equation}}
\def\ee{\end{equation}}

\def\bea{\begin{eqnarray}}
\def\eea{\end{eqnarray}}

\title{Non-minimal Kinetic coupling to gravity and accelerated expansion}
\author{L.N. Granda\thanks{ngranda@univalle.edu.co}\\ {\small\it Departamento de Fisica, Universidad del Valle}} 
\date{}
\begin{document}
\maketitle

\begin{abstract}
\noindent We study a scalar field with kinetic term coupled to itself and to the curvature, as a source of dark energy, and analyze the role of this new coupling in the accelerated expansion at large times. In the case of scalar field dominance, the scalar field and potential giving rise to power-law expansion are found in some cases, and a dynamical equation of state is calculated for a given solution of the field equations. A behavior very close to that of the cosmological constant was found.\\ 

\end{abstract}

\section{Introduction}
\noindent 
Recent astrophysical data from distant Ia supernovae observations \cite{SN},\cite{riess} show that the current Universe is not only expanding, it is accelerating due to some kind of  negative-pressure form of matter called dark energy. This dark energy may consist of cosmological constant, conventionally associated with the energy of the vacuum  or alternatively, could came from a dynamical varying scalar field at late times which also account for the missing energy density in the universe. That kind of scalar fields are allowed from several theories in particle physics and in multidimensional gravity, like Kaluza-Klein theory, String Theory\cite{green} or Supergravity, in which the scalar field appears in a natural way.
The cosmological solutions using scalar fields have been a subject of intensive study in the last years in the usual model called quintessence \cite{RP}, \cite{wett}, \cite{shani}, \cite{stein}, \cite{copeland97}, \cite{stein1}, \cite{stein2}, where the stability of scaling solutions was studied (i.e. solutions where the scalar energy density scale in the same way like a barotropic fluid) for a field evolving in accordance with an exponential potential and a power law potential, which provide a late time inflation. More exotic approaches to the problem of dark energy using scalar fields are related with K-essence models, based on scalar field with non-standard kinetic term \cite{stein3},\cite{chiba}; string theory fundamental scalars known as tachyons \cite{pad}; scalar field with negative kinetic energy, which provides a solution known as phantom dark energy \cite{caldwell} (see \cite{copeland} for a review). The non-minimal coupling between the quintessence field and curvature, as an explanation of the accelerated expansion, has been considered among others, in refs. \cite{chiba1} \cite{uzan},\cite{wands}. This coupling has also been studied in the context of the inflationary cosmology \cite{amendola1}, \cite{maeda},  \cite{turner}, \cite{kasper}, \cite{easson1}. An inflationary model including the DBI term and non-minimally coupled scalar has been proposed in \cite{easson2,easson3}\\
\noindent In this paper we consider the possibility of an explicit coupling between the scalar field, the kinetic term and the curvature, as a source of dark energy, and analyze the role of this new coupling in an evolution scenario with late-time accelerated expansion. The basic motivation for studying such theories is related with the fact that they appear as lowenergy limit of several higher dimensional theories, e.g. superstring theory \cite{green}, and provide a possible approach to quantum gravity from a perturbative point of view. They appear as part of the Weyl anomaly in $N=4$ conformal supergravity \cite{tseytlin, odintsov2}. A model with non-minimal derivative couplings was proposed in \cite{amendola2}, \cite{capozziello1}, \cite{capozziello2} in the context of inflationary cosmology, and in \cite{caldwell1} a derivative coupling to Ricci tensor has been considered to study cosmological restrictions on the coupling parameter, and the role of this coupling during inflation. Some asymptotical solutions for a non-minimal kinetic coupling to scalar and Ricci curvatures were found in \cite{sushkov}. 
Non-minimal coupling of arbitrary function $f(R)$ with matter Lagrangian (including the kinetic scalar term) has been introduced in \cite{sergei},\cite{allemandi}. Such a model was proposed to describe the dark energy, late-time universe acceleration. When $f(R)$ represents the power law function, such a model (which maybe considered as string-inspired theory) maybe proposed for dynamical resolution of cosmological constant problem as it has been suggested in \cite{sergei1},\cite{sergei2}. In the present work we consider this function $f(R)$ as linear in $R$ but we 
generalize the model permitting extra $R_{\mu\nu}$ coupling with kinetic-like scalar term. In addition, we keep the free kinetic term for scalar.\\
\noindent We propose a model of dark energy in which the scalar-field kinetic term is non-minimally coupled to itself and to the scalar and Ricci curvatures. The considered self coupling have the advantage of leaving the coupling constants without dimension. Some asymptotic solutions for the scalar field are proposed, and the scalar-field potential is found. The power-law expansion is also considered, and the expressions for the scalar-field and potential are found. In the last section we present some conclusions.

\section{Field Equations}

Let us start with the following  action:

\be\label{eq1}
\begin{aligned}
S=&\int d^{4}x\sqrt{-g}\Big[\frac{1}{16\pi G} R-\frac{1}{2}\partial_{\mu}\phi\partial^{\mu}\phi-\frac{1}{2} \xi R \left(\frac{1}{\phi^2}\partial_{\mu}\phi\partial^{\mu}\phi\right) -\\ 
&\frac{1}{2} \eta R_{\mu\nu}\left(\frac{1}{\phi^2}\partial^{\mu}\phi\partial^{\nu}\phi\right) - V(\phi)\Big] + S_m.
\end{aligned}
\ee

\noindent where $S_m$ is the dark matter action which describes a fluid with barotropic equation of state. The presence of the $1/\phi^2$ coefficient in each interacting term, guarantees that the coupling constants $\xi$ and $\eta$ are dimensionless. Taking the variation of action \ref{eq1} with respect to the metric, we obtain a general expression of the form 
\be\label{eq2}
R_{\mu\nu}-\frac{1}{2}g_{\mu\nu}R=\kappa^2\left[T_{\mu\nu}^m+T_{\mu\nu}\right]
\ee
where $\kappa^2=8\pi G$, $T_{\mu\nu}^m$ is the usual energy-momentum tensor for matter component, and the tensor $T_{\mu\nu}$ represents the variation of the terms which depend on the scalar field $\phi$ and can be written as
\be\label{eq3}
T_{\mu\nu}=T_{\mu\nu}^{\phi}+T_{\mu\nu}^{\xi}+T_{\mu\nu}^{\eta}
\ee
where $T_{\mu\nu}^{\phi}$, $T_{\mu\nu}^{\xi}$, $T_{\mu\nu}^{\eta}$ correspond to the variations of the minimally coupled terms, the $\xi$ and the $\eta$ couplings respectively. Those variations are given by
\be\label{eq4}
T_{\mu\nu}^{\phi}=\nabla_{\mu}\phi\nabla_{\nu}\phi-\frac{1}{2}g_{\mu\nu}\nabla_{\lambda}\phi\nabla^{\lambda}\phi
-g_{\mu\nu}V(\phi)
\ee
\be\label{eq5}
\begin{aligned}
T_{\mu\nu}^{\xi}=&\xi\Big[\left(R_{\mu\nu}-\frac{1}{2}g_{\mu\nu}R\right)\left(\frac{1}{\phi^2}\nabla_{\lambda}\phi\nabla^{\lambda}\phi\right)+g_{\mu\nu}\nabla_{\lambda}\nabla^{\lambda}\left(\frac{1}{\phi^2}\nabla_{\gamma}\phi\nabla^{\gamma}\phi\right)\\
&-\frac{1}{2}(\nabla_{\mu}\nabla_{\nu}+\nabla_{\nu}\nabla_{\mu})\left(\frac{1}{\phi^2}\nabla_{\lambda}\phi\nabla^{\lambda}\phi\right)+R\left(\frac{1}{\phi^2}\nabla_{\mu}\phi\nabla_{\nu}\phi\right)\Big]
\end{aligned}
\ee
\be\label{eq6}
\begin{aligned}
T_{\mu\nu}^{\eta}=&\eta\Big[\frac{1}{\phi^2}\left(R_{\mu\lambda}\nabla^{\lambda}\phi\nabla_{\nu}\phi+R_{\nu\lambda}\nabla^{\lambda}\phi\nabla_{\mu}\phi\right)-\frac{1}{2}g_{\mu\nu}R_{\lambda\gamma}\left(\frac{1}{\phi^2}\nabla^{\lambda}\phi\nabla^{\gamma}\phi\right)\\
&-\frac{1}{2}\left(\nabla_{\lambda}\nabla_{\mu}\left(\frac{1}{\phi^2}\nabla^{\lambda}\phi\nabla_{\nu}\phi\right)+\nabla_{\lambda}\nabla_{\nu}\left(\frac{1}{\phi^2}\nabla^{\lambda}\phi\nabla_{\mu}\phi\right)\right)\\
&+\frac{1}{2}\nabla_{\lambda}\nabla^{\lambda}\left(\frac{1}{\phi^2}\nabla_{\mu}\phi\nabla_{\nu}\phi\right)+\frac{1}{2}g_{\mu\nu}\nabla_{\lambda}\nabla_{\gamma}\left(\frac{1}{\phi^2}\nabla^{\lambda}\phi\nabla^{\gamma}\phi\right)\Big]
\end{aligned}
\ee
Due to the interacting terms this expressions are defined in the Jordan frame and do not correspond to the energy-momentum tensors as defined in the Einstein frame. We consider that in action (\ref{eq1}) matter is universally coupled only to $g_{\mu\nu}$, and thus all experimental data will have their usual interpretation in this frame. Variating with respect to the scalar field gives rise to the equation of motion
\be\label{eq7}
\begin{aligned}
&-\frac{1}{\sqrt{-g}}\partial_{\mu}\left[\sqrt{-g}\left(\xi R\frac{\partial^{\mu}\phi}{\phi^2}+\eta R^{\mu\nu}\frac{\partial_{\nu}\phi}{\phi^2}+\partial^{\mu}\phi\right)\right]+\frac{dV}{d\phi}-\\
&\frac{1}{\phi^3}\left(\xi R\partial_{\mu}\phi\partial^{\mu}\phi+\eta R_{\mu\nu}\partial^{\mu}\phi\partial^{\nu}\phi\right)=0
\end{aligned}
\ee
Assuming the spatially-flat FriedmannRobertsonWalker (FRW) metric,
\be\label{eq8}
ds^2=-dt^2+a(t)^2\left(dr^2+r^2d\Omega^2\right)
\ee
using (\ref{eq4}-\ref{eq6}), we can write the $(00)$ and $(11)$ components of the tensor $T_{\mu\nu}$  (with the Hubble parameter $H$ and for homogeneous time-depending scalar field) as follows
\be\label{eq9}
\begin{aligned}
T_{00}=&\frac{1}{2}\dot{\phi}^2+V(\phi)+3\xi\left[(3H^2+2\dot{H})\frac{\dot{\phi}^2}{\phi^2}-2H\frac{\dot{\phi}\ddot{\phi}}{\phi^2}+2H\frac{\dot{\phi}^3}{\phi^3}\right]+\\
&3\eta\left[\dot{H}\frac{\dot{\phi}^2}{\phi^2}-H\frac{\dot{\phi}\ddot{\phi}}{\phi^2}+H\frac{\dot{\phi}^3}{\phi^3}\right]
\end{aligned}
\ee
and
\be\label{eq10}
\begin{aligned}
\frac{T_{11}}{a^2}=&\frac{1}{2}\dot{\phi}^2-V(\phi)+\xi\left[(3H^2+2\dot{H})\frac{\dot{\phi}^2}{\phi^2}+4H\left(\frac{\dot{\phi}\ddot{\phi}}{\phi^2}-\frac{\dot{\phi}^3}{\phi^3}\right)+2\left(\frac{\ddot{\phi}^2}{\phi^2}+\frac{\dot{\phi}\dddot{\phi}}{\phi^2}-5\frac{\dot{\phi}^2\ddot{\phi}}{\phi^3}+3\frac{\dot{\phi}^4}{\phi^4}\right)\right]+\\
&\eta\left[(3H^2+2\dot{H})\frac{\dot{\phi}^2}{\phi^2}+4H\left(\frac{\dot{\phi}\ddot{\phi}}{\phi^2}-\frac{\dot{\phi}^3}{\phi^3}\right)+\frac{\ddot{\phi}^2}{\phi^2}+\frac{\dot{\phi}\dddot{\phi}}{\phi^2}-5\frac{\dot{\phi}^2\ddot{\phi}}{\phi^3}+3\frac{\dot{\phi}^4}{\phi^4}\right]
\end{aligned}
\ee
which can be interpreted as the effective density and pressure respectively. The equation of motion \ref{eq7} takes the form
\be\label{eq11}
\begin{aligned}
&\ddot{\phi}+3H\dot{\phi}+\frac{dV}{d\phi}+3(4\xi+\eta)H^2\left(\frac{\ddot{\phi}}{\phi^2}-\frac{\dot{\phi}^2}{\phi^3}\right)+3(2\xi+\eta)\dot{H}\left(\frac{\ddot{\phi}}{\phi^2}-\frac{\dot{\phi}^2}{\phi^3}\right)\\
&+9(4\xi+\eta)H^3\frac{\dot{\phi}}{\phi^2}+3(14\xi+5\eta)H\dot{H}\frac{\dot{\phi}}{\phi^2}+3(2\xi+\eta)\ddot{H}\frac{\dot{\phi}}{\phi^2}=0
\end{aligned}
\ee
where the first three terms correspond to the minimally coupled scalar field.\\
\noindent In order to simplify this equations and find some simple solutions we impose the restriction on $\xi$ and $\eta$ given by $\eta+2\xi=0$. This restriction simplifies the modified Friedmann equation for the $(00)$-component (see Eq. (\ref{eq2})), which takes the form
\be\label{eq12}
H^2=\frac{\kappa^2}{3}\left(\frac{1}{2}\dot{\phi}^2+V(\phi)+9\xi H^2\frac{\dot{\phi}^2}{\phi^2}\right)
\ee
where we have replaced $\eta=-2\xi$, and from now on we will consider only the contribution of the scalar field (i.e. $T_{\mu\nu}^m=0$). For the $(11)$-component of Eq. (\ref{eq2}) it is obtained
\be\label{eq13}
-2\dot{H}-3H^2=\kappa^2\left[\frac{1}{2}\dot{\phi}^2-V(\phi)-\xi\left(3H^2+2\dot{H}\right)\frac{\dot{\phi}^2}{\phi^2}-4\xi H\left(\frac{\dot{\phi}\ddot{\phi}}{\phi^2}-\frac{\dot{\phi}^3}{\phi^3}\right)\right]
\ee
and the equation of motion also reduces to
\be\label{eq14}
\ddot{\phi}+3H\dot{\phi}+\frac{dV}{d\phi}+6\xi H^2\left(\frac{\ddot{\phi}}{\phi^2}-\frac{\dot{\phi}^2}{\phi^3}\right)
+18\xi H^3\frac{\dot{\phi}}{\phi^2}+12\xi H\dot{H}\frac{\dot{\phi}}{\phi^2}=0
\ee

Next we try to find cosmological implications of the Eqs. (\ref{eq12}) and (\ref{eq14}), and look for solutions giving rise to accelerated expansion and acceptable behavior of the equation of state parameter (EoS). First note that if we consider the asymptotic solution $\phi=\phi_0=const.$, then from Eqs. (\ref{eq12}) and (\ref{eq14}) it follows that $V=V_0=const$ and $H=H_0=\kappa\sqrt{V_0/3}$ and we get an asymptotic de Sitter behavior.


\section{Power-law solutions}
In what follows we will consider the particular case of the model (\ref{eq1}) without free kinetic term. In this case we are in the frames of above  models (\cite{sergei,allemandi,sergei1,sergei2}) which correspond to pure dark energy scalar sector. This also applies if the the slow-roll condition $\dot{\phi}^2<<V(\phi)$ is considered, which can take place in the contexts of dark energy
or inflationary cosmology. Let us consider the effects of this new kinetic coupling in the cosmological dynamics, in the case of scalar field dominance (we further neglect any background radiation or matter contribution).

\noindent{\bf Derivative coupling}\\

Initially we will look for effects strictly related to the derivative coupling.
Let us then, begin with the simple case of only non-minimal kinetic coupling, without potential term. In this case the Eq. (\ref{eq12}) reduces to
\be\label{eq15}
3\xi\kappa^2\frac{\dot{\phi}^2}{\phi^2}=1
\ee
which has a solution
\be\label{eq16}
\phi=\phi_0 e^{\pm \frac{t}{\sqrt{3\xi}\kappa}}
\ee
and using this solution, the Eq. (\ref{eq14}) takes the simple form 
\be\label{eq16a}
3H^2+2\dot{H}=0
\ee
and therefore $H=2/(3t)$, which gives a power-law corresponding to pressureless matter dominance. Note that substituting back to Lagrangian the solution (\ref{eq15}), (\ref{eq16}), we get kind of non-covariant modified gravity (for general review , see (\cite{sergei3})).\\
Going back to Eqs. (\ref{eq9}-\ref{eq11}) with arbitrary $\xi$ and $\eta$ and without free kinetic and potential terms, the Friedmann equation and the equation of motion from Eqs. (\ref{eq9}) and (\ref{eq11}) take the form
\be\label{eq17}
H^2=\xi\left((3H^2+2\dot{H})\chi-H\dot{\chi}\right)+\eta\left(\dot{H}\chi-\frac{1}{2}H\dot{\chi}\right)
\ee
\be\label{eq18}
(4\xi+\eta)H^2\dot{\chi}+(2\xi+\eta)\dot{H}\dot{\chi}
+2\left(3(4\xi+\eta)H^3+(14\xi+5\eta)H\dot{H}+(2\xi+\eta)\ddot{H}\right)\chi=0
\ee
where we have multiplied the Eq. (\ref{eq14}) by $\dot{\phi}$, and introduced the new variable $\chi=\kappa^2\frac{\dot{\phi}^2}{\phi^2}$. Let's propose now a power law solution $H=p/t$, and replace in Eq. (\ref{eq18}) to obtain
\be\label{eq19}
\left[(4\xi+\eta)p-(2\xi+\eta)\right]\dot{\chi}+2\left[3(4\xi+\eta)p^2-(14\xi+5\eta)p+2(2\xi+\eta)\right]\chi=0
\ee
The solution to this equation is of the form
\be\label{eq20}
\chi=t^{\alpha},\,\,\,\, \alpha=\frac{3(4\xi+\eta)p^2-(14\xi+5\eta)p+2(2\xi+\eta)}{2\xi+\eta-(4\xi+\eta)p}
\ee
where we considered the condition $\chi_0=1$ at $t_0=1$. Replacing this solution in Eq. (\ref{eq17}), we obtain the following restrictions on the parameters
\be\nonumber
p=\frac{2\xi+\eta}{3\xi-1}+\frac{(2\xi+\eta)\alpha}{2(3\xi-1)},
\ee
\be\label{eq21}
3(4\xi+\eta)p^2-(14\xi+5\eta)p+2(2\xi+\eta)=0
\ee
the second one comes from the condition $\alpha=0$, giving rise to the simple restriction for accelerated expansion ($p>1$): $\xi>1/3, \xi-\eta<1$ or $\xi<1/3, \xi-\eta>1$. Replacing $p$ from the first condition in the second one (using $\alpha=0$), gives an equation for $\xi$ and $\eta$, which has the three following solutions 
\be\label{eq22}
\eta=-2/3,\,\,\,\,\,\, \eta=-2\xi,\,\,\,\,\,\,\, \eta=-\xi-1
\ee
the solution $\eta=-2/3$ can not give rise to accelerated expansion, the second solution gives $p=0$, and the third solution gives accelerated expansion on the line $\eta=-\xi-1$ in the interval $0<\xi<1/3$.\\
From Eq. (\ref{eq12}), the effective gravitational coupling can be expressed as
\be\label{eq22a}
\kappa_{eff}^2=8\pi G_{eff}= \kappa^2(1-3\xi\kappa^2\frac{\dot{\phi}^2}{\phi^2})^{-1}
\ee
from the solutions Eqs. (\ref{eq16}), (\ref{eq20}) ($\alpha=0$), it follows that $\dot{G_{eff}}/G_{eff}=0$, satisfying the restrictions on the time variation of the gravitational coupling.\\

\noindent{\bf Derivative coupling with potential}\\

Let's now consider the scalar field potential. We will look for the shape of the potential $V(\phi)$ corresponding to the asymptotic behavior giving rise to accelerated expansion.
In this case, adopting the same variable $\chi=\kappa^2\frac{\dot{\phi}^2}{\phi^2}$, the Eq. (\ref{eq12}) takes the form

\be\label{eq23}
H^2=\frac{\kappa^2}{3}\frac{V(\phi)}{1-3\xi\chi}
\ee
and multiplying by $\kappa^2\dot{\phi}$ the Eq. of motion (\ref{eq14}), it's obtained
\be\label{eq24}
\kappa^2\frac{dV}{dt}+3\xi H^2\frac{d\chi}{dt}+18\xi H^3\chi+12\xi H\dot{H}\chi=0
\ee
where the first two terms of Eq. (\ref{eq14}) have been dropped due to the absence of the free kinetic term.
Taking the derivative of the potential in Eq. (\ref{eq23}), and replacing in Eq. (\ref{eq24}), we obtain
\be\label{eq25}
\xi H^2\frac{d\chi}{dt}+\xi H\dot{H}\chi-3\xi H^3\chi-H\dot{H}=0
\ee
looking for power-law solutions, we replace $H=p/t$ in Eq. (\ref{eq25}), obtaining
\be\label{eq26}
\xi t\frac{d\chi}{dt}-\xi(1+3p)\chi+1=0
\ee
considering the particular solution $\chi=\frac{1}{\xi(1+3p)}$, gives the scalar field $\phi$ 
\be\label{eq26a}
\phi=\phi_0\exp\left(-\frac{t}{\kappa\sqrt{\xi(1+3p)}}\right)
\ee
\noindent Then, the scalar field potential giving rise to power-law expansion is obtained from Eq. (\ref{eq23}) by replacing $H$ and $\phi$
\be\label{eq27}
V(\phi)=\frac{3p^2}{\kappa^4\xi}\left(\frac{3p-2}{(3p+1)^2}\right)\frac{1}{\log^2{\phi/\phi_0}}
\ee
\noindent Note that we have not restriction on $p$, so for $p>1$ it follows the accelerated expansion.\\
\noindent The general solution of Eq. (\ref{eq26}) is given by
\be\label{eq28}
\chi=t^{1+3p}+\frac{1}{\xi(1+3p)}
\ee
which gives for the scalar field 
\be\label{eq29}
\begin{aligned}
&\log{\frac{\phi}{\phi_0}}=\frac{2t\left(1+\xi(1+3p)t^{1+3p}\right)^{1/2}}{3(1+p)\left(\xi(1+3p)\right)^{1/2}}\\
&+\frac{t\left(\xi(1+3p)\right)^{1/2}}{3\xi(1+p)}\text{Hypergeometric2F1}\left[\frac{1}{1+3p},\frac{1}{2},1+\frac{1}{1+3p},-\xi(1+3p)t^{1+3p}\right]
\end{aligned}
\ee
In this case the scalar field potential can not be solved explicitly in terms of the field $\phi$, but the time dependence as follows from (\ref{eq28}) and $H$, is given by 
\be\label{eq30}
V(t)=\frac{3p^2}{\kappa^2}\left(\frac{3p-2}{3p+1}t^{-2}-3\xi t^{3p-1}\right)
\ee
which gives $V(\phi)$ through Eq.(\ref{eq29}). For such field and potential, it is possible to have power-law expansion with $p>1$. Note that this potential contains two asymptotic behaviors: in the limit $t\rightarrow0$ the $t^{-2}$ term dominates in Eq. (\ref{eq30}) (for $p>1/3$) and the potential behaves like potential (\ref{eq27}), as can be seen by the Taylor expansion of Eq. (\ref{eq29}) for a given values of $p>0$ and $\xi>0$ (in this case $\log(\phi/\phi_0)\sim t+ \text{higher orders}$). In the limit $t\rightarrow\infty$, the term with $t^{3p-1}$ dominates in the potential (\ref{eq30})(for $p>1/3$), and replacing this approximation in Eq. (\ref{eq23})
\be\label{eq30a}
p^2t^{-2}=\frac{-3\xi p^2t^{3p-1}}{1-3\xi t^{3p+1}-\frac{3}{3p+1}}
\ee
gives $p=2/3$. Thus dropping the $t^{-2}$ term in (\ref{eq30}) at large times, the power-law solution $H=p/t$ still valid, but has the restriction $p=2/3$, which gives an EoS typical of pressureless dark matter, coinciding with the previous result obtained in (\ref{eq16a}).\\
\noindent Alternatively we can propose a late time asymptotical solution of the form $\kappa^2\frac{\dot{\phi}^2}{\phi^2}=c^2$ with $c$ constant, and replacing in Eq. (\ref{eq23}) we find that the denominator becomes constant and $H^2$ takes the simple form
\be\label{eq31}
H(t)^2=\frac{\kappa^2}{3}\bar{V}(t),\,\,\,\,\,\,  \bar{V}(t)=\frac{V(t)}{1-3\xi c^2}
\ee
and replacing in Eq. (\ref{eq14}), the following equation for $\bar{V}$ is obtained
\be\label{eq32}
(1-\xi c^2)\frac{d\bar{V}}{dt}+2\sqrt{3}\xi\kappa c^2\bar{V}^{3/2}=0
\ee
which has a simple solution 
\be\label{eq33}
V=\left(\frac{\eta}{t}\right)^2,\,\,\,\,\,\,  \eta=\frac{1-\xi c^2}{\sqrt{3}\xi\kappa c^2}
\ee
giving power law solution $H=p/t$, with $p=(1-\xi c^2)/\sqrt{3}\xi c^2$. In this case the condition for accelerated expansion $p>1$, is satisfied for $\xi c^2<1/(\sqrt{3}+1)$.
The time variation of the effective gravitational coupling for the particular solution (\ref{eq26a}) is zero, as in the previous cases, and for the solution (\ref{eq28}) it is obtained
\be\label{eq33a}
\frac{\dot{G_{eff}}}{G_{eff}}\Big|_{t_0=1}=\sigma H_0,\,\,\,\,\,\,\,\,  \sigma=\frac{3\xi(1+3p)^2}{p(3p-2-3\xi(1+3p))}
\ee
where we used $d\chi/dt|_{t_0}=(1+3p)/p  H_0$. Taking $H_0\approx(6-7)\times10^{-11} yr^{-1}$, $\sigma$ should be of the order of $10^{-2}$ or less, to satisfy the current constraints on the variation of the gravitational coupling \cite{uzan1}. Taking for instance $p=2$, $\xi$ should be $\xi\lesssim 5\times10^{-4}$.

\section{Dynamically varying equation of state}
In this case we propose an asymptotical solution for the scalar field at late times, and try to reconstruct the Hubble parameter and the potential, giving rise to dynamically changing EOS parameter, evolving according to the current observations. 
Let us write the equations (\ref{eq12}) and (\ref{eq14}) in terms of the variable $x=\log a$. The Eq. (\ref{eq12}) without free kinetic term becomes
\be\label{eq34}
H^2=\frac{\kappa^2}{3}\left(V(\phi)+9\xi H^4\frac{{\phi'}^2}{\phi^2}\right)
\ee
and the Eq. of motion (\ref{eq14}) takes the form
\be\label{eq35}
\frac{1}{\phi'}\frac{dV}{dx}+6\xi H^2\left(\frac{\phi''}{\phi^2}-\frac{\phi'^2}{\phi^2}+3\frac{\phi'}{\phi^2}\right)+9\xi\frac{\phi'}{\phi^2}\frac{dH^2}{dx}=0
\ee
where $\phi'=d\phi/dx$ and we have used $dV/d\phi=1/\phi'dV/dx$.\\
\noindent Resolving the Eq. (\ref{eq34}) with respect to $H^2$ one obtains
\be\label{eq36}
H^2=\frac{1}{6\xi\kappa^2\frac{\phi'^2}{\phi^2}}\left(1\pm\sqrt{1-4\xi\kappa^4 V\frac{\phi'^2}{\phi^2}}\right)
\ee
here we can take both signs of the root, using  $\xi>0$ for the $(+)$ root and $\xi<0$ for the negative one, in order to keep $H^2$ positive (the choice of the sign does'n influence the Eq. (\ref{eq35}), as $V$ in Eq. (\ref{eq36}) is under the square root). Considering the particular form of the scalar field $\phi=\phi_0 e^{-\alpha x}$ and replacing in Eq. (\ref{eq36}) we get
\be\label{eq37}
H^2=\frac{\kappa^2}{3}\tilde{V}(x),\,\,\,\,\,\,  \tilde{V}(x)=\frac{1}{2\xi\alpha^2\kappa^4}\left(1\pm\sqrt{1-4\xi\alpha^2\kappa^4 V}\right)
\ee
setting $\kappa^2=1$ and replacing Eq. (\ref{eq37}) in Eq. (\ref{eq35}), the following equation for the new potential $\tilde{V}(x)$ is obtained
\be\label{eq38}
\left(\xi\alpha^2\tilde{V}(x)-1\right)\frac{d\tilde{V}(x)}{dx}-2\xi\alpha^2\tilde{V}^2(x)=0
\ee
Integrating this equation we obtain the solution for $\tilde{V}$ as follows
\be\label{eq39}
\tilde{V}(x)=-\left(\xi\alpha^2\text{ProductLog}\left[-\frac{1}{\xi\alpha^2V_0}e^{-2x-1/(\xi\alpha^2V_0)}\right]\right)^{-1}
\ee 
using this solution in Eq. (\ref{eq37}) (with $\kappa^2=1$), and changing to the redshift variable $z$, the Hubble parameter is obtained as follows  
\be\label{eq40}   
H^2(z)=-\left(3\xi\alpha^2\text{ProductLog}\left[-\frac{1}{\xi\alpha^2V_0}e^{-1/(\xi\alpha^2V_0)}(1+z)^2\right]\right)^{-1}
\ee
From the properties of the $\text{ProductLog}$ function follows that for $\xi>0$ (here $V_0>0$), the argument of the $\text{ProductLog}$ function will be always negative, and if the argument is less than $-1/e$ then the function $\text{ProductLog}$ becomes complex, so taking into account the sign of the argument in Eq. (\ref{eq40}), the region of interest in our case is restricted by 
\be 
\frac{1}{\xi\alpha^2V_0}e^{-1/(\xi\alpha^2V_0)}(1+z)^2<\frac{1}{e}
\ee
but for given values of $V_0>0, \xi>0$ and $\alpha$, the equation
\be
\frac{1}{\xi\alpha^2V_0}e^{-1/(\xi\alpha^2V_0)}(1+z)^2=\frac{1}{e}
\ee
has always two real $z$ roots. Thus, for a given values of the parameters $V_0>0, \xi>0$ and $\alpha$ there will be always two real values of z delimiting a forbidden region. Therefore we exclude positive values of $\xi$ in this analysis. By other hand, for negative $\xi$, $H^2$ will be always real positive as follows from ($\ref{eq40}$) and the properties of the $\text{ProductLog}$ function. From the Eq. (\ref{eq37}) and the expression for $H^2$, it follows that the effective dark energy density given by $\rho_{eff}=3H^2/\kappa^2=\tilde{V}$ tends to zero at high redshifts, validating the assumption that the dark energy was negligible at early times, when the matter contribution was relevant, but presents a Big Rip singularity at the future revealing phantom behavior. 

\noindent We can now evaluate the effective equation of state
\be\label{eq41}
w_{eff}=-1-\frac{1}{3}\frac{1}{H^2}\frac{dH^2}{dx}
\ee
and using Eqs. (\ref{eq37}) and \ref{eq39}, we can write $w_{eff}$ in therms of the redshift as
\be\label{eq42}
w_{eff}=-1-\frac{2}{3\left(\text{ProductLog}\left[-\frac{1}{\xi\alpha^2V_0}e^{-1/(\xi\alpha^2V_0)}(1+z)^2\right]+1\right)}
\ee
An interesting fact of this expression is that the $\text{ProductLog}$ function varies very slowly with $z$ for a given combination of $\xi, \alpha, V_0$. Fig.1 shows the behavior of the effective equation of state for two different combinations of the parameters, represented as $\gamma=\xi\alpha^2 V_0$.
\begin{center}
\includegraphics [scale=0.7]{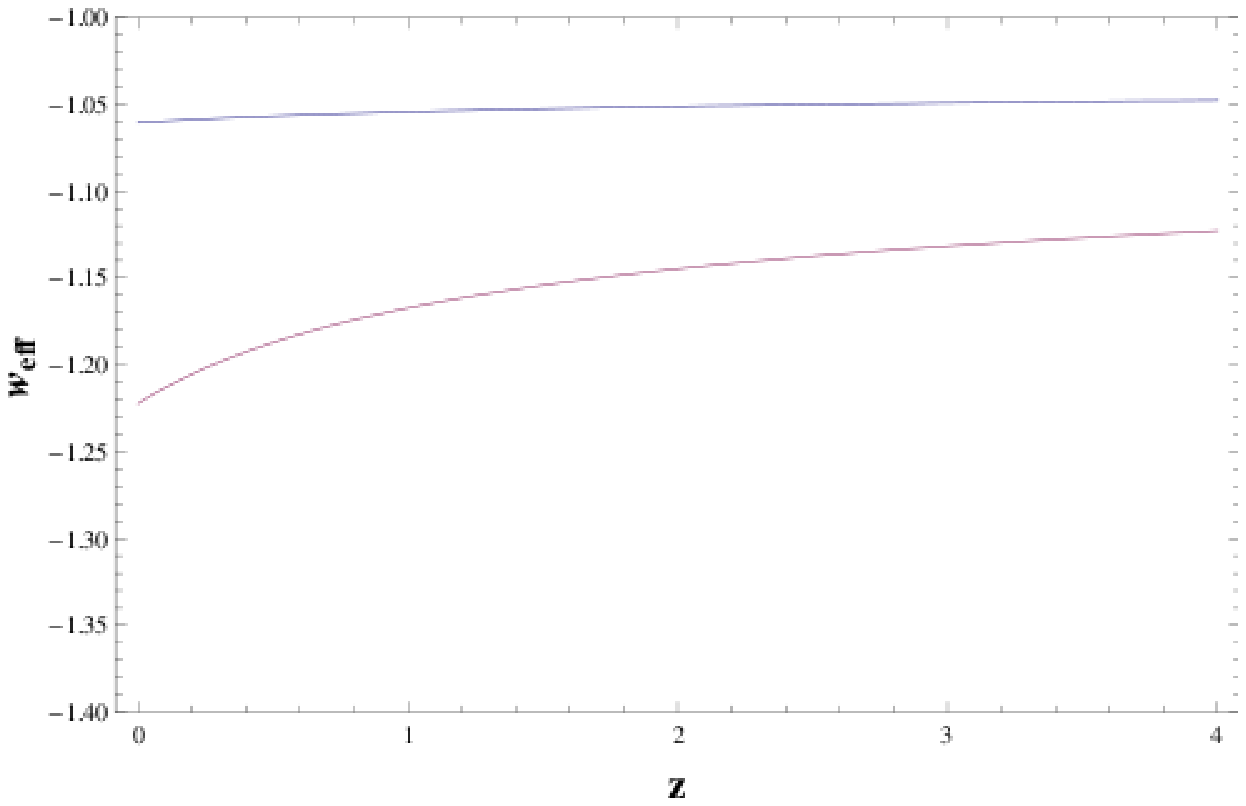}
\end{center}
\begin{center}
{Fig. 1 \it The dark energy effective equation of state parameter $w_{eff}$ versus redshift, for two combinations of the parameters $\gamma=-0.1$ (top), and $\gamma=-0.5$ (bottom), where $\gamma=\xi\alpha^2 V_0$} 
\end{center}
Note that for the product of parameters $\gamma=\xi\alpha^2 V_0=-0.1$, the effective EoS parameter varies from $w_{eff}\approx-1.06$ at the present ($z=0$) to $-1$ at $z\rightarrow\infty$, showing very slowly variation with time, and maintaining very close to $-1$. This behavior is very close to that of the cosmological constant. In this case the time variation of the effective gravitational coupling for the scalar field $\phi=\phi_0 e^{-\alpha x}$, can be written as
\be\label{eq42a}
\frac{\dot{G_{eff}}}{G_{eff}}\Big|_{t_0=1}=\sigma H_0,\,\,\,\,\,\,\,\,  \sigma=\frac{6\xi\alpha^2\kappa^2\dot{H}_0}{1-3\xi\alpha^2\kappa^2H_0^2}
\ee
considering $\dot{H}_0\lesssim H_0^2$, and due to the extremelly small value of $\kappa^2H_0^2=M_{Pl}^{-2}H_0^2$, is always possible to find a suitable combination of $\xi\alpha^2$ that satisfy the current constraints on $\dot{G_{eff}}/G_{eff}$ \cite{uzan1}.
\section{Discussion}
We have considered a new model of non-minimally kinetic coupled scalar field to explain the nature of the dark energy. The kinetic scalar term is not only coupled to itself, but to the curvature, giving rise to interesting cosmological consecuences. In the absence of a potential, this model presents power-law solutions. First we note that the strictly non-minimal kinetic coupling produces the effect of pressureless dark matter (\ref{eq16a}), which is the behavior of the matter at early stage of the evolution, when the curvature could play an important role. Continuing with the strictly non-minimal kinetic coupling, but in the more general case, without the restriction on the coupling constants $\xi$ and $\eta$ (i.e. $\eta\neq-2\xi$), the model has late time power-law solutions leading to accelerated expansion (see Eq. (\ref{eq20})). In the presence of a potential $V(\phi)$, the model presents power-law solutions giving rise to accelerated expansion, as showed in Eqs. (\ref{eq26a})-(\ref{eq30}) and (\ref{eq33}). Additionally, the potential (\ref{eq30}) may be interpreted in two asymptotical cases, corresponding to early and late time behavior. Note that we have only considered power law solutions in absence of matter. It is thus expected that including a matter term completely changes the evolution of the scale factor, possibly giving a past deceleration followed by the present acceleration, according to the current observations.\\
\noindent Looking for solutions in the presence of potential, but giving rise to dynamical effective EoS (not of the power-law type), we proposed a solution for the scalar field of the form $\phi=\phi_0 e^{-\alpha x}$, and found the potential as presented in Eq. (\ref{eq39}). According to this result, the model presents phantom behavior, but with an EoS very close to that of the cosmological constant, as seen from Fig. 1. Note from Eq. (\ref{eq42}) and Fig.1, that when the combination $\gamma=-\xi\alpha^2 V_0\rightarrow-0$, the effective EoS parameter $w_{eff}\rightarrow-1$, behaving very close to the cosmological constant. The effective EoS parameter varies from $-1$ in the far past $z\rightarrow\infty$ to $-5/3$ in the future ($z=-1$), but is worth to note that Eq. (\ref{eq42}) allows any current value of $w_{eff0}$ in the limits $-5/3<w_{eff0}<-1$, and this value is closer to $-1$ for small values of $\gamma$. For instance, for $\gamma=-0.1$, the effective EoS parameter changes from $-1.06$ to $-1$ in a practically infinite time, giving very exact description of the cosmological constant. The study of solutions to this model with matter contribution will be done elsewhere . Note that in models with non-minimal coupling to f(R), we can consider also $f(G)$ (where $G$ is Gauss-Bonnet combination) as function multiplied to kinetic scalar term. Then it would of interest to study generalization of model (\ref{eq1}) where instead of $R$ we use function $G$ (see (\cite{sergei4})).\\
\noindent In conclusion, non-minimal kinetic coupling provides a new framework to study solutions to the dark energy problem. This model provides late-time accelerated expansion even without potential as shown by Eqs. (\ref{eq20}-\ref{eq22}). This indicates that the derivative couplings provide an effective scalar density and pressure which might play an important role in the explanation of the dark energy or cosmological constant problem.

\section*{Acknowledgments}
This work was supported by Universidad del Valle, project N CI-7796.

\end{document}